\documentclass {entcs}
\usepackage [utf8] {inputenc}
\usepackage [T1] {fontenc}
\usepackage {url}
\usepackage{listings}
\usepackage{xcolor}

\definecolor{dkblue}{rgb}{0,0.1,0.5}
\definecolor{lightblue}{rgb}{0,0.5,0.5}
\definecolor{dkgreen}{rgb}{0,0.6,0}
\definecolor{dk2green}{rgb}{0,0.2,0}
\definecolor{dkviolet}{rgb}{0.6,0,0.8}

\lstdefinelanguage{Coq}{
mathescape=true,
texcl=false,
basicstyle=\ttfamily\footnotesize\bf,
morekeywords=[1]{
Section, Module, End, Require, Import, Export,
Variable, Variables, Parameter, Parameters, Axiom, Hypothesis, Hypotheses,
Notation, Local, Tactic, Reserved, Scope, Open, Close, Bind, Delimit,
Definition, Let, Ltac, Fixpoint, CoFixpoint, Add, Morphism, Relation, Function,
Implicit, Arguments, Set, Unset, Contextual, Strict, Prenex, Implicits,
Inductive, CoInductive, Record, Structure, Canonical, Coercion,
Context, Class, Global, Instance, Program, Infix,
Theorem, Lemma, Corollary, Proposition, Fact, Remark, Example,
Proof, Goal, Save, Qed, Defined, Hint, Resolve, Rewrite, View,
Search, Show, Print, Printing, All, Graph, Projections, inside, outside},
keywordstyle=[1]{\ttfamily\color{dkviolet}},
morekeywords=[2]{forall, exists, exists2, fun, fix, cofix, struct,
      match, with, end, as, in, return, let, if, is, then, else,
      for, of, nosimpl, where},
keywordstyle=[2]{\ttfamily\color{dkgreen}},
morekeywords=[3]{Type, Prop},
keywordstyle=[3]{\ttfamily\color{lightblue}},
morekeywords=[4]{
         pose, set, move, case, elim, apply, clear,
         hnf, intro, intros, generalize, rename, pattern, after,
         destruct, induction, using, refine, inversion, injection,
         rewrite, congr, unlock, compute, field, fourier,
         replace, fold, unfold, change, cutrewrite, simpl,
         have, suff, wlog, suffices, without, loss, nat_norm,
         assert, cut, trivial, revert, bool_congr, nat_congr,
         symmetry, transitivity, auto, split, left, right, autorewrite},
keywordstyle=[4]{\ttfamily\color{dkblue}},
morekeywords=[5]{
         by, done, exact, tauto, romega, omega,
         assumption, solve, contradiction, discriminate},
keywordstyle=[5]{\ttfamily\color{red}},
identifierstyle={\ttfamily\color{black}},
keywordstyle=[6]{\ttfamily\color{dkpink}},
stringstyle=\sffamily,
commentstyle={\ttfamily\scriptsize},
comment=[s]{(*}{*)},
showstringspaces=false,
morestring=[b]",
morestring=[d]’,
tabsize=3,
extendedchars=false,
sensitive=true,
breaklines=true,
}

\begin {document}
\begin{frontmatter}
\title {Formalization of simplification for context-free grammars}
\author{Marcus V. M. Ramos	
\thanksref{myemail}
}
\address{Centro de Informática\\UFPE\\Recife, Brazil} 
\thanks[myemail]{Email:\href{mailto:mvmr@cin.ufpe.br} {\texttt{\normalshape mvmr@cin.ufpe.br}}}	

\author{Ruy J. G. B. de Queiroz
\thanksref{coemail}
}
\address{Centro de Informática\\UFPE\\Recife, Brazil} 
\thanks[coemail]{Email:\href{mailto:ruy@cin.ufpe.br} {\texttt{\normalshape ruy@cin.ufpe.br}}}	
\begin {abstract}
Context-free grammar simplification is a subject of high importance in computer language processing technology as well as in formal language theory. This paper presents a formalization, using the Coq proof assistant, of the fact that general context-free grammars generate languages that can be also generated by simpler and equivalent context-free grammars. Namely, useless symbol elimination, inaccessible symbol elimination, unit rules elimination and empty rules elimination operations were described and proven correct with respect to the preservation of the language generated by the original grammar.
\end {abstract}

\end{frontmatter}

\section {Introduction}
\label {sec-intro}
The formalization of context-free language theory is key to the certification of compilers and programs, as well as to the development of new languages and tools for certified programming. The results presented is this paper are part of an ongoing work that intends to formalize parts of the context-free language theory in the Coq proof assistant. The initial results comprised the formalization of closure properties for context-free grammars, namely union, concatenation and Kleene star \cite {ramos-2014}.

In order to follow this paper, the reader is required to have basic knowledge of Coq and of context-free language theory. For the beginner, the recommended starting points for Coq are the book by Bertot \cite {bertot-2004}, the online book by Pierce \cite {pierce} and a few tutorials available on \cite {coq-site}. Detailed information on the Coq proof assistant, as well as on the syntax and semantics of the following definitions and statements, is available in \cite {coq-2012}. Background on context-free language theory can be found in \cite {sudkamp-2006}, \cite {hopcroft-1979} or \cite {ramos-2009}, among others.

The objective of this work is to formalize a substantial part of context-free language theory in the Coq proof assistant, making it possible to reason about it in a fully checked environment, with all the related advantages. Initially, however, the focus has been restricted to context-free grammars and associated results. Pushdown automata and their relation to context-free grammars will be considered in the future.

When the work is complete, it should be useful for a few different purposes. Among them, to make available a complete and mathematically precise description of the behavior of the objects of context-free language theory. Second, to offer fully checked and mechanized demonstrations of its main results. Third, to provide a library with basic and fundamental lemmas and theorems about context-free grammars and derivations that can be used as a starting point to prove new theorems and increase the amount of formalization for context-free language theory. Fourth, to allow for the certified and efficient implementation of its relevant algorithms in a programming language. Fifth, to permit the experimentation in an educational environment in the form of a tool set, in a laboratory where further practical observations and developments can be done, for the benefit of students, teachers, professionals and researchers.

The general idea of formalizing context-free language theory in the Coq proof assistant is discussed in Section \ref {sec-basic}. The methodology used is briefly reviewed in Section \ref {sec-methodology}. Specific results related to the formalization of grammar simplification are presented in Section \ref {sec-simplification}. The plan for the rest of this research is presented in Section \ref {sec-further}, and Section \ref {sec-related} considers related work by various other researchers.

The results reported in this paper are related to the elimination of symbols (terminals and non-terminals) in context-free grammars that do not contribute to the language being generated, and also to the elimination of unit and empty rules, in order to shorten the derivation of the sentences of the language.

All the definitions and proof scripts presented in this paper were written in plain Coq and are available for download at: \\
\url {https://github.com/mvmramos/simplification} \\

\section {Basic Definitions}
\label {sec-basic}
Context-free grammars were represented in Coq very closely to the usual algebraic definition $G=(V,\Sigma,P,S)$, where $V$ is the vocabulary of $G$ (it includes all non-terminal and terminal symbols), $\Sigma$ is the set of terminal symbols (used in the construction of the sentences of the language generated by the grammar), $N=V\setminus\Sigma$ is the set of non-terminal symbols (representing different sentence abstractions), $P$ is the set of rules and $S \in N$ is the start symbol (also called initial or root symbol). Rules have the form $\alpha \rightarrow \beta$, with $\alpha \in N$ and $\beta \in V^*$.

Basic definitions in Coq are presented below. The $N$ and $\Sigma$ sets are represented separately from $G$ (respectively by types \texttt {non\_terminal} and \texttt {terminal}). The disjoint union of the types \texttt {non\_terminal} and \texttt {terminal} is represented by the symbol \texttt {+}. Notations \texttt {sf} (sentential form) and \texttt {sentence} represent lists, possibly empty, of respectively terminal and non-terminal symbols and terminal only symbols.

\begin{lstlisting}[mathescape]
Variables non_terminal terminal: Type.
Notation sf := (list (non_terminal + terminal)).
Notation sentence := (list terminal).
Notation nlist:= (list non_terminal).
\end{lstlisting}

The record representation \texttt {cfg} has been used for $G$. The definition states that \texttt {cfg} is a new type and contains three components. The first is the \texttt {start\_symbol} of the grammar (a non-terminal symbol) and the second is \texttt {rules}, that represent the rules of the grammar. Rules are propositions (represented in Coq by \texttt {Prop}) that take as arguments a non-terminal symbol and a (possibly empty) list of non-terminal and terminal symbols (corresponding, respectively, to the left and right-hand side of a rule). 

The predicate \texttt {rules\_finite\_def} assures that the set of rules of the grammar is finite by proving that the length of right-hand side of every rule is equal or less than a given value, and also that both left and right-hand side of the rules are built from finite sets of, respectively, non-terminal and terminal symbols (represented here by lists).

\begin{lstlisting}
Definition rules_finite_def (ss: non_terminal) 
                            (rules: non_terminal -> sf -> Prop) 
                            (n: nat) 
                            (ntl: list non_terminal)
                            (tl: list terminal) :=
In ss ntl /\
(forall left: non_terminal,
 forall right: list (non_terminal + terminal),
 rules left right ->
 length right <= n /\
 In left ntl /\
 (forall s : non_terminal, In (inl s) right -> In s ntl) /\
 (forall s : terminal, In (inr s) right -> In s tl)).

Record cfg: Type:= {
start_symbol: non_terminal;
rules: non_terminal -> sf -> Prop;
rules_finite: exists n: nat,
              exists ntl: nlist,
              exists tl: tlist,
              rules_finite_def start_symbol rules n ntl tl }.
\end{lstlisting}

The decision of representing rules as propositions has the consequence that it will prevent executable code to be extracted from the formalization. It would surely be desirable to be able to obtain certified algorithms for, in the present case, the simplification of context-free grammars. The alternative then would be to represent rules as a member of type \texttt {list (non\_terminal * sf)} instead. This, however, would have changed the whole declarative approach of the present work into a more computational one, by creating functions that manipulate grammars that have the desired properties. The purely logical approach, thus, was considered more appealing and selected as the choice for the present formalization. Anyway, it does not affect the objectives listed in Section \ref {sec-intro} and can be adapted in the future in order to allow for code extraction, although this should demand a considerable effort in the creation and proof of program-related scripts.

The example below represents grammar that generates language $a^*b$: $$G=(\{S',A,B,a,b\},\{a,b\},\{S' \rightarrow aS', S' \rightarrow b\},S')$$ The following are the definitions used to represent $G$ in Coq (as \texttt {g}):

\begin{lstlisting}
Inductive non_terminal: Type:=
| S'
| A
| B.

Inductive terminal: Type:=
| a
| b.

Inductive rs: non_terminal -> sf -> Prop:=
  r1: rs S' [inr a; inl S']
| r2: rs S' [inr b].

Definition g: cfg _ _:= {|
start_symbol:= S'; 
rules:= rs;
rules_finite:= rs_finite |}.
\end{lstlisting}

The term \texttt {rs\_finite} (the proof that the set of rules of \texttt {g} is finite) is not presented here, but can be easily constructed and is available from the link provided in Section \ref {sec-intro}. 

Another fundamental concept used in this formalization is the idea of \emph {derivation}: a grammar \texttt{g} \emph {derives} a string \texttt {s2} from a string \texttt {s1} if there exists a series of rules in \texttt {g} that, when applied to \texttt {s1}, eventually result in \texttt {s2}. An inductive predicate definition of this concept in Coq (\texttt {derives}) uses two constructors.

\begin{lstlisting}
Inductive derives (g: cfg): sf g -> sf g -> Prop :=
| derives_refl: forall s: sf g,
                derives g s s
| derives_step: forall s1 s2 s3: sf g,
                forall left: non_terminal g,
                forall right: sf g,
                derives g s1 (s2 ++ inl left :: s3) ->
                rules g left right ->
                derives g s1 (s2 ++ right ++ s3).
\end{lstlisting}

The constructors of this definition (\texttt {derives\_refl} and \texttt {derives\_step}) are the axioms of our theory. Constructor \texttt {derives\_refl} asserts that every sentential form \texttt {s} can be derived from \texttt {s} itself. Constructor \texttt {derives\_step} states that if a sentential form that contains the left-hand side of a rule is derived by a grammar, then the grammar derives the sentential form with the left-hand side replaced by the right-hand side of the same rule. This case corresponds to the application of a rule in a direct derivation step.

A grammar \texttt {generates} a string if this string can be derived from its root symbol. Finally, a grammar \texttt {produces} a sentence if it can be generated from its root symbol.

\begin{lstlisting}
Definition generates (g: cfg) (s: sf): Prop:=
derives g [inl (start_symbol g)] s.

Definition produces (g: cfg) (s: sentence): Prop:=
generates g (map terminal_lift s).
\end{lstlisting} 

Function \texttt {terminal\_lift} converts a terminal symbol into an ordered pair of type \texttt {(non\_terminal + terminal)}. With these definitions, it has been possible to prove various lemmas about grammars and derivations, and also operations on grammars, all of which were useful when proving the main theorems of this article. 

As an example, the lemma that states that $G$ produces the string $aab$ (that is, that $aab \in L(G)$) is represented as:

\begin{lstlisting}
Lemma G_produces_aab:
produces G [a; a; b].
\end{lstlisting} 

The proof of this lemma can be easily constructed and relates directly to the derivations in $S \Rightarrow aS \Rightarrow aaS \Rightarrow aab$, however in reverse order because of the way that \texttt {derives} is defined.

\section {Methodology}
\label {sec-methodology}

This formalization is about the definition of a new contex-free grammar from a previous one, such that it (i) both grammars generate the same language and (ii) the new grammar is free of a certain kind of symbols or rules. For all the four cases considered, the following common approach has been adopted:

\begin {enumerate}
\item Depending on the case, inductively define a new type of non-terminal symbols; this will be important, for example, when we want to guarantee that the start symbol of the grammar does not appear in the right-hand side of any rule or when we have to construct new non-terminal symbols from the existing ones;
\item Inductively define the rules of the new grammar, in a way that allows the construction of the proofs that the resulting grammar has the required properties; these new rules will likely make use of the new non-terminal symbols described above;
\item Define the new grammar by using the new non-terminal symbols and the new rules; define the new start symbol (which might be a new non-terminal symbol or an existing one) and build a proof of the finiteness of the set of rules for this new grammar;
\item State and prove all the lemmas and theorems that will assert that the newly defined grammar has the desired properties.
\end {enumerate}

In the following section, this approach will be explored with further detail for each main result achieved in this work.

\section {Simplification}
\label {sec-simplification}

The definition of a context-free grammar allows for the inclusion of symbols and rules that might not contribute to the language being generated. Also, context-free grammars might also contain sets of rules that can be substituted by equivalent smaller and simpler sets of rules. Unit rules, for example, do not expand sentential forms (instead, they just rename the symbols in them) and empty rules can cause them to contract. Although the appropriate use of these features can be important for human communication in some situations, this is not the general case, since it leads to grammars that have more symbols and rules than necessary, making difficult its comprehension and manipulation. Thus, simplification is an important operation on context-free grammars.

Let $G$ be a context-free grammar, $L(G)$ the language generated by this grammar and $\epsilon$ the empty string. Different authors use different terminology when presenting simplification results for context-free grammars. In what follows, we adopt the terminology and definitions of \cite {sudkamp-2006}. 

Context-free grammar simplification comprises four kinds of objects, whose definitions and results are presented below:

\begin {enumerate}
\item \label {empty}
An \emph {empty rule} $r \in P$ is a rule whose right-hand side $\beta$ is empty (e.g. $X \rightarrow \epsilon$). We formalize that for all $G$, there exists $G'$ such that $L(G)=L(G')$ and $G'$ has no empty rules, except for a single rule $S \rightarrow \epsilon$ if $\epsilon \in L(G)$; in this case, $S$ (the initial symbol of $G'$) does not appear in the right-hand side of any rule in $G'$;

\item \label {unit} 
A \emph {unit rule} $r \in P$ is a rule whose right-hand side $\beta$ contains a single non-terminal symbol (e.g. $X \rightarrow Y$). We formalize that for all $G$, there exists $G'$ such that $L(G)=L(G')$ and $G'$ has no unit rules;

\item \label {useless} 
$s \in V$ is \emph {useful} (\cite {sudkamp-2006}, p. 116) if it is possible to derive a string of terminal symbols from it using the rules of the grammar. Otherwise $s$ is called a \emph {useless symbol}. A useful symbol $s$ is one such that $s \Rightarrow^* \omega$, with $\omega \in \Sigma^*$. Naturally, this definition concerns mainly non-terminals, as terminals are trivially useful. We formalize that, for all $G$ such that $L(G) \neq \emptyset$, there exists $G'$ such that $L(G)=L(G')$ and $G'$ has no useless symbols;

\item \label {inaccessible} 
$s \in V$ is \emph {accessible} (\cite {sudkamp-2006}, p. 119) if it is part of at least one string generated from the root symbol of the grammar. Otherwise, it is called an \emph {inaccessible symbol}. An accessible symbol $s$ is one such that $S \Rightarrow^* \alpha s\beta$, with $\alpha, \beta \in V^*$. We formalize that for all $G$, there exists $G'$ such that $L(G)=L(G')$ and $G'$ has no inaccessible symbols.
\end {enumerate}

Finally, we formalize a unification result: that for all $G$, if $G$ is non-empty, then there exists $G'$ such that $L(G)=L(G')$ and $G'$ has no empty rules (except for one, if $G$ generates the empty string), no unit rules, no useless symbols and no inaccessible symbols.

In all these four cases and five grammars that are discussed next (namely \texttt {g\_emp}, \texttt {g\_emp'}, \texttt {g\_unit}, \texttt {g\_use} and \texttt {g\_acc}), the proof of the predicate \texttt {rules\_finite} is based on the proof of the correspondent predicate for the argument grammar. Thus, all new grammars satisfy the \texttt {cfg} specification and are finite as well.

\subsection {Empty rules}
Result (\ref {empty}) is achieved in two steps. First, the idea of a \emph {nullable} symbol was represented by the definition \texttt {empty}: 

\begin{lstlisting}
Definition empty 
(g: cfg terminal _) (s: non_terminal + terminal): Prop:=
derives g [s] [].
\end{lstlisting}

Notation \texttt {sf'} represents a sentential form built with symbols from \texttt {non\_terminal'} and \texttt {terminal}. Definition \texttt {symbol\_lift} maps a pair of type \texttt {(non\_terminal + terminal)} into a pair of type \texttt {(non\_terminal' + terminal)} by replacing each \texttt {non\_terminal} with the corresponding \texttt {non\_terminal'}:

\begin{lstlisting}
Inductive non_terminal': Type:=
| Lift_nt: non_terminal -> non_terminal'
| New_ss. 

Notation sf' := (list (non_terminal' + terminal)).

Definition symbol_lift 
(s: non_terminal + terminal): non_terminal' + terminal:=
match s with
| inr t => inr t
| inl n => inl (Lift_nt n)
end.
\end{lstlisting}

With these, a new grammar \texttt {g\_emp g} has been created, such that the language generated by it matches the language generated by the original grammar (\texttt {g}), except for the empty string. Predicate \texttt {g\_emp\_rules} states that every non-empty rule of \texttt {g} is also a rule of \texttt {g\_emp g}, and also adds new rules to \texttt {g\_emp g} where every possible combination of nullable non-terminal symbols that appears in the right-hand side of a rule of \texttt {g} is removed, as long as the resulting right-hand side is not empty. Finally, it adds a rule that maps a new symbol, the start symbol of the new grammar (\texttt {New\_ss}), to the start symbol of the original grammar. For this reason, the new type \texttt {non\_terminal'} has been defined. The motivation for introducing a new start symbol at this point is to be able to prove that the start symbol does not appear in the right-hand side of any rule of the new grammar, a result that will be important in future developments.

\begin{lstlisting}
Inductive g_emp_rules (g: cfg _ _): non_terminal' -> sf' -> Prop :=
| Lift_direct : 
       forall left: non_terminal,
       forall right: sf,
       right <> [] -> rules g left right ->
       g_emp_rules g (Lift_nt left) (map symbol_lift right)
| Lift_indirect:
       forall left: non_terminal,
       forall right: sf,
       g_emp_rules g (Lift_nt left) (map symbol_lift right)->
       forall s1 s2: sf, 
       forall s: non_terminal,
       right = s1 ++ (inl s) :: s2 ->
       empty g (inl s) ->
       s1 ++ s2 <> [] ->
       g_emp_rules g (Lift_nt left) (map symbol_lift (s1 ++ s2))
| Lift_start_emp: 
       g_emp_rules g New_ss [inl (Lift_nt (start_symbol g))]. 
	   
Definition g_emp (g: cfg non_terminal terminal): 
cfg non_terminal' terminal := {|
start_symbol:= New_ss;
rules:= g_emp_rules g;
rules_finite:= g_emp_finite g |}.
\end{lstlisting}

Suppose, for example, that $S, A, B, C$ are non-terminals, of which $A, B$ and $C$ are nullable, $a, b$ and $c$ are terminals and $X \rightarrow aAbBcC$ is a rule of \texttt {g}. Then, the above definitions assert that $X \rightarrow aAbBcC$ is a rule of \texttt {g\_emp g}, and also:

\begin {itemize}
\item $X \rightarrow aAbBc$;
\item $X \rightarrow abBcC$;
\item $X \rightarrow aAbcC$;
\item $X \rightarrow aAbc$;
\item $X \rightarrow abBc$;
\item $X \rightarrow abcC$;
\item $X \rightarrow abc$.
\end {itemize}

Observe that grammar \texttt {g\_emp g} does not generate the empty string. The second step, thus, was to define \texttt {g\_emp' g}, such that \texttt {g\_emp' g} generates the empty string if \texttt {g} generates the empty string. This was done by stating that every rule from \texttt {g\_emp g} is also a rule of \texttt {g\_emp' g} and also by adding a new rule that allow \texttt {g\_emp' g} to generate the empty string directly if necessary. 

\begin{lstlisting}
Inductive g_emp'_rules (g: cfg _ _): 
non_terminal' non_terminal -> sf' -> Prop :=
| Lift_all:
       forall left: non_terminal' _,
       forall right: sf',
       rules (g_emp g) left right ->
       g_emp'_rules g left right
| Lift_empty:
       empty g (inl (start_symbol g)) -> 
       g_emp'_rules g (start_symbol (g_emp g)) [].

Definition g_emp' (g: cfg non_terminal terminal): 
cfg (non_terminal' _) terminal := {|
start_symbol:= New_ss _;
rules:= g_emp'_rules g;
rules_finite:= g_emp'_finite g |}.
\end{lstlisting}

Note that the generation of the empty string by \texttt {g\_emp' g} depends on \texttt {g} generating the empty string. 

The proof of the correctness of these definitions is achieved through the following theorem:

\begin{lstlisting}
Theorem g_emp'_correct: 
forall g: cfg non_terminal terminal,
g_equiv (g_emp' g) g /\
(generates_empty g -> has_one_empty_rule (g_emp' g)) /\ 
(~ generates_empty g -> has_no_empty_rules (g_emp' g)) /\
start_symbol_not_in_rhs (g_emp' g).
\end{lstlisting}

Four auxiliary predicates have been used in this statement: \texttt {g\_equiv} for two context-free grammars that generate the same language, \texttt {generates\_empty} for a grammar whose language includes the empty string, \texttt {has\_one\_empty\_rule} for a grammar that has an empty rule whose left-hand side is the initial symbol, and all other rules are not empty and \texttt {has\_no\_empty\_rules} for a grammar that has no empty rules at all.

The definition of \texttt {g\_equiv} is straightforward:

\begin{lstlisting}
Variables non_terminal non_terminal' terminal: Type.

Definition g_equiv (g1: cfg non_terminal terminal) 
                   (g2: cfg non_terminal' terminal): Prop:=
forall s: sentence,
produces g1 s <-> produces g2 s.
\end{lstlisting}

When applied to the previous theorem, it translates into:

\begin{lstlisting}
forall s: sentence,
produces (g_emp' g) s <-> produces g s.
\end{lstlisting}

For the \texttt {->} part, the strategy adopted is to prove that for every rule $left \rightarrow_{g\_emp'} right$ of (\texttt {g\_emp' g}), either $left \rightarrow_{g} right$ is a rule of \texttt {g} or $left \Rightarrow^*_{g} right$ in \texttt {g}. For the \texttt {<-} part, the strategy is a more complicated one, and involves induction over the number of derivation steps in \texttt {g}.

\subsection {Unit rules}

For result (\ref {unit}), definition \texttt {unit} expresses the relation between any two non-terminal symbols $X$ and $Y$, and is true when $X \Rightarrow^* Y$.

\begin{lstlisting}
Inductive unit (g: cfg terminal non_terminal) (a: non_terminal): 
non_terminal -> Prop:=
| unit_rule: forall (b: non_terminal),
             rules g a [inl b] -> unit g a b
| unit_trans: forall b c: non_terminal,
              unit g a b ->
              unit g b c ->
              unit g a c.
\end{lstlisting}

Grammar \texttt {g\_unit g} represents the grammar whose unit rules have been substituted by equivalent ones. The idea is that \texttt {g\_unit g} has all non-unit rules of \texttt {g}, plus new rules that are created by anticipating the possible application of unit rules in \texttt {g}, as informed by \texttt {g\_unit}.

\begin{lstlisting}
Inductive g_unit_rules (g: cfg _ _): non_terminal -> sf -> Prop :=
| Lift_direct' : 
       forall left: non_terminal,
       forall right: sf,
       (forall r: non_terminal,
       right <> [inl r]) -> rules g left right ->
       g_unit_rules g left right
| Lift_indirect':
       forall a b: non_terminal,
       unit g a b ->
       forall right: sf,
       rules g b right ->  
       (forall c: non_terminal,
       right <> [inl c]) -> 
       g_unit_rules g a right.

Definition g_unit (g: cfg _ _): cfg _ _ := {|
start_symbol:= start_symbol g;
rules:= g_unit_rules g;
rules_finite:= g_unit_finite g |}.
\end{lstlisting}

Finally, the correcteness of \texttt {g\_unit} comes from the following theorem:

\begin{lstlisting}
Theorem g_unit_correct: 
forall g: cfg _ _,
g_equiv (g_unit g) g /\
has_no_unit_rules (g_unit g).
\end{lstlisting}

The predicate \texttt {has\_no\_unit\_rules} states that the argument grammar has no unit rules at all.

Similar to the previous case, for the \texttt {->} part of the \texttt {g\_equiv (g\_unit g) g} proof, the strategy adopted is to prove that for every rule $left \rightarrow_{g\_unit} right$ of (\texttt {g\_unit g}), either $left \rightarrow_{g} right$ is a rule of \texttt {g} or $left \Rightarrow^*_{g} right$ in \texttt {g}. For the \texttt {<-} part, the strategy is also a more complicated one, and involves induction over a predicate that is isomorphic to \emph {derives} (\emph {derives3}), but generates the sentence directly without considering the application of a sequence of rules, which allows one to abstract the application of unit rules in \texttt {g}.

\subsection {Useless symbols}
For result (\ref {useless}), the idea of a useful symbol is captured by the definition \texttt {useful}.

\begin{lstlisting}
Definition useful (g: cfg _ _) (s: non_terminal + terminal): Prop:=
match s with
| inr t => True
| inl n => exists s: sentence, derives g [inl n] (map term_lift s)
end.
\end{lstlisting}

The removal of useless symbols comprises, first, the identification of useless symbols in the grammar and, second, the elimination of the rules that use them. Definition \texttt {g\_use\_rules} selects, from the original grammar, only the rules that do not contain useless symbols. The new grammar, without useless symbols, can then be defined as in \texttt {g\_use}.

\begin{lstlisting}
Inductive g_use_rules (g: cfg): non_terminal -> sf -> Prop :=
| Lift_use : forall left: non_terminal,
             forall right: sf,
             rules g left right ->
             useful g (inl left) ->
             (forall s: non_terminal + terminal, In s right -> 
             useful g s) -> g_use_rules g left right.

Definition g_use (g: cfg _ _): cfg _ _:= {|
start_symbol:= start_symbol g;
rules:= g_use_rules g;
rules_finite:= g_use_finite g |}.			 
\end{lstlisting}

The \texttt {g\_use} definition, of course, can only be used if the language generated by the original grammar is not empty, that is, if the root symbol of the original grammar is useful. If it were useless then it would be impossible to assign a root to the grammar and the language would be empty. The correctness of the useless symbol elimination operation can be certified by proving theorem \texttt {g\_use\_correct}, which states that every context-free grammar whose root symbol is useful generates a language that can also be generated by an equivalent context-free grammar whose symbols are all useful.

\begin{lstlisting}
Theorem g_use_correct: 
forall g: cfg _ _,
non_empty g ->
g_equiv (g_use g) g /\
has_no_useless_symbols (g_use g). 
\end{lstlisting}

The predicates \texttt {non\_empty}, and \texttt {has\_no\_useless\_symbols} used above assert, respectively, that grammar \texttt {g} generates a language that contains at least one string (which in turn may or may not be empty) and the grammar has no useless symbols at all.

The \texttt {->} part of the \texttt {g\_equiv} proof is straightforward, since every rule of \texttt {g\_use} is also a rule of \texttt {g}. For the converse, it is necessary to show that every symbol used in a derivation of \texttt {g} is useful, and thus the rules used in this derivation also appear in \texttt {g\_use}.

\subsection {Inaccessible symbols}

Result (\ref {inaccessible}) is similar to the previous case, and definition \texttt {accessible} has been used to represent accessible symbols in context-free grammars.

\begin{lstlisting}
Definition accessible 
(g: cfg _ _) (s: non_terminal + terminal): Prop:=
exists s1 s2: sf, derives g [inl (start_symbol g)] (s1++s::s2).
\end{lstlisting}

Definition \texttt {g\_acc\_rules} selects, from the original grammar, only the rules that do not contain inaccessible symbols. Definition \texttt {g\_acc} represents a grammar whose inaccessible symbols have been removed.

\begin{lstlisting}
Inductive g_acc_rules (g: cfg): non_terminal -> sf -> Prop :=
| Lift_acc : forall left: non_terminal,
             forall right: sf,
             rules g left right -> accessible g (inl left) -> 
             g_acc_rules g left right.

Definition g_acc (g: cfg _ _): cfg _ _ := {|
start_symbol:= start_symbol g;
rules:= g_acc_rules g;
rules_finite:= g_acc_finite g |}.
\end{lstlisting}

The correctness of the inaccessible symbol elimination operation can be certified by proving theorem \texttt {g\_acc\_correct}, which states that every context-free grammar generates a language that can also be generated by an equivalent context-free grammar whose symbols are all accessible.

\begin{lstlisting}
Theorem g_acc_correct: 
forall g: cfg _ _,
g_equiv (g_acc g) g /\
has_no_inaccessible_symbols (g_acc g).  
\end{lstlisting}

In a way similar to \texttt {has\_no\_useless\_symbols}, the absence of inaccessible symbols in a grammar is expressed by predicate \texttt {has\_no\_inaccessible\_symbols} used above.

Similar to the previous case, the \texttt {->} part of the \texttt {g\_equiv} proof is also straightforward, since every rule of \texttt {g\_acc} is also a rule of \texttt {g}. For the converse, it is necessary to show that every symbol used in a derivation of \texttt {g} is accessible, and thus the rules used in this derivation also appear in \texttt {g\_acc}.

\subsection {Unification}

If one wants to obtain a new grammar simultaneously free of empty and unit rules, and of useless and inaccessible symbols, it is not enough to consider the previous independent results. On the other hand, it is necessary to establish a suitable order to apply these simplifications, in order to guarantee that the final result satisfies all desired conditions. Then, it is necessary to prove that the claims do hold.

For the order, we should start with (i) the elimination of empty rules, followed by (ii) the elimination of unit rules. The reason for this is that (i) might introduce new unit rules in the grammar, and (ii) will surely not introduce empty rules, as long as original grammar is free of them (except for $S \rightarrow \epsilon$, in which case $S$, the initial symbol of the grammar, must not appear in the right-hand side of any rule). Then, elimination of useless and inaccessible symbols (in either order) is the right thing to do, since they only remove rules from the original grammar (which is specially important because they do not introduce new empty or unit rules).

The formalization of this result is captured in the following theorem, which represents the main result of this work:

\begin{lstlisting}
Theorem g_simpl_exists_v1:
forall g: cfg non_terminal terminal,
 non_empty g ->
 exists g': cfg (non_terminal' non_terminal) terminal,
 g_equiv g' g /\
 has_no_inaccessible_symbols g' /\
 has_no_useless_symbols g' /\
(generates_empty g -> has_one_empty_rule g') /\ 
(~ generates_empty g -> has_no_empty_rules g') /\
 has_no_unit_rules g' /\
 start_symbol_not_in_rhs g'.
\end{lstlisting}

Hypothesis \texttt {non\_empty g} is necessary in order to allow the elimination of useless symbols. The predicate \texttt {start\_symbol\_not\_in\_rhs} states that the start symbol does not appear in the right-hand side of any rule of the argument grammar.

The proof of \texttt {g\_simpl\_exists\_v1} demands auxiliary lemmas to prove that the characteristics of the initial transformations are preserved by the following ones. For example, unit rules elimination, useless symbol elimination and inaccessible symbol elimination operations preserve the characteristics of the empty rules elimination operation.

The proofs of all lemmas and theorems presented in this article have been formalized in Coq and comprises approximately 10,000 lines of scripts. This number can be explained for the following reasons:

\begin {enumerate}
\item The style adopted for writing the scripts: for the sake of clarity, each tactic is placed in its own line, despite the possibility of combining several tactics in the same line. Also, bullets (for structuring the code) were used as much as possible and the sequence tactical (using the semicolon symbol) was avoided at all. This duplicates parts of the code but has the advantage of keeping the static structure of the script related to its dynamic behaviour, which favors legibility and maintenance.
\item The formalization includes not only the main theorems described here, but also an extensive library of other fundamental and auxiliary lemmas on context-free grammars and derivations, which have been used to obtain the main results presented here, were used in the previously obtained results and will be used in future developments.
\end {enumerate}

\section {Further Work}
\label {sec-further}

Current work has focussed on the representation of context-free grammars, context-free derivations, the formalization of grammar simplification strategies and the certification of their correctness. It represents an important step towards the formalization of context-free language theory, and adds to the previous results on the formalization of closure properties for context-free grammars (\cite {ramos-2014}). 

The next steps of this formalization work are:

\begin {enumerate}
\item Describe Chomsky normal form for context-free grammars and prove its existence for any context-free grammar that satisfies the required conditions;
\item Obtain a formal proof of the Pumping Lemma for context-free languages. 
\end {enumerate}

The second objective relies on the first one, while the first depends directly on the results presented here.

\section {Related Work}
\label {sec-related}
Language and automata theory has been subject of formalization since the mid-1980s, when Kreitz used the Nuprl proof assistant to prove results about  deterministic finite automata and the pumping lemma for regular languages \cite {kreitz-1986}. Since then, the theory of regular languages has been formalized partially by different researchers using different proof assistants (see \cite {constable-1997}, \cite {kaloper-1996}, \cite {filliatre-1997a}, \cite {briais-2008}, \cite {miyamoto}, \cite {moreira-2009}, \cite {almeida-2010a}, \cite {almeida-2010b}, \cite {moreira-2012} \cite {braibant-2010}, \cite {braibant-2012}, \cite {asperti-2012b}, \cite {coquand-2011}, \cite {krauss-2012} and \cite {wu-2011}). The most recent and complete formalization, however, is the work by Jan-Oliver Kaiser \cite {doczkal-2013}, which used Coq and the SSReflect extension to prove the main results of regular language theory.

Context-free language theory has not been formalized to the same extent so far, and the results were obtained with a diversity of proof assistants, including Coq, HOL4 and Agda. Most of the effort start in 2010 and has been devoted to the certification and validation of parser generators. Examples of this are the works of Koprowski and Binsztok (using Coq, \cite {koprowski-2010}), Ridge (using HOL4, \cite {ridge-2011}), Jourdan, Pottier and Leroy (using Coq, \cite {jourdan-2012}) and, more recently, Firsov and Uustalu (in Coq, \cite {firsov-2014}). 

On the more theoretical side, on which the present work should be considered, Norrish and Barthwal (using HOL4, \cite {barthwal-norrish-2010a}, \cite {barthwal-norrish-2010b}, \cite {barthwal-norrish-2013}), published on general context-free language theory formalization, including the existence of normal forms for grammars, pushdown automata and closure properties. Recently, Firsov and Uustalu proved the existence of a Chomsky Normal Form grammar for every general context-free grammar (using Agda, \cite {firsov-2015}).

It can thus be noted that apparently no formalization has been done in Coq so far for results not related directly to parsing and parser verification, and that this constitutes an important motivation for the present work, mainly due to the increasing usage and importance of Coq in different areas and communities. Specifically, the formalization done by Norrish and Barthwal in HOL4 is quite comprehensive and extends our work with the Greibach Normal Form and pushdown automata and its relation to context-free grammars. It does not include, however, a proof of either the decidability of the membership problem or the Pumping Lemma for context-free languages, which are objectives of the present work. The formalization by Firsov and Uustalu in Agda comprises basically the existence of a Chomsky Normal Form, and formalizes the elimination of empty and unit rules, but not elimination of useless and inaccessible symbols.

When it comes to computability theory and Turing machines related classes of languages, formalization has been approached by Asperti and Ricciotti (Matita, \cite {asperti-2012b}), Xu, Zhang and Urban (Isabelle/HOL, \cite {xu-2013}) and Norrish (HOL4, \cite {norrish-2011}).

\section {Conclusions}

The present paper reports an ongoing effort towards formalizing the classical context-free language theory, initially based only on context-free grammars, in the Coq proof assistant. All important objects have been formalized and different simplification strategies on grammars have been implemented. Proofs of their correctness were successfully constructed.

Building up on the previous formalization of closure properties for context-free grammars \cite {ramos-2014}, the present results create a comfortable situation in order to pursue the formalization of normal forms for context-free grammars, the next step of this work.

The authors acknowledge the fruitful discussions with Nelma Moreira (Departamento de Ciência de Computadores da Faculdade de Ciências da Universidade do Porto, Portugal), José Carlos Bacelar Almeida (Departamento de Informática da Universidade do Minho, Portugal) and Arhur Azevedo de Amorim (University of Pennsylvania) as well as their contributions to this work. Also, we are grateful to the anonymous reviewers who provided useful criticisms and insights, and contributed to a better presentation of this work.

\bibliographystyle {entcs}
\bibliography {article}

\begin{thebibliography}{10}
\expandafter\ifx\csname url\endcsname\relax
  \def\url#1{\texttt{#1}}\fi
\expandafter\ifx\csname urlprefix\endcsname\relax\def\urlprefix{URL }\fi
\newcommand{\enquote}[1]{``#1''}

\bibitem{almeida-2010b}
Almeida, J. C.~B., N.~Moreira, D.~Pereira and S.~M. de~Sousa, \emph{Partial
  derivative automata formalized in {C}oq}, in: \emph{Proceedings of the 15th
  International Conference on Implementation and Application of Automata},
  CIAA'10 (2011), pp. 59--68.
\newline\urlprefix\url{http://dl.acm.org/citation.cfm?id=1964285.1964292}

\bibitem{almeida-2010a}
Almeida, M., N.~Moreira and R.~Reis, \emph{Testing the equivalence of regular
  languages}, Journal of Automata, Languages and Combinatorics \textbf{15}
  (2010), pp.~7--25.

\bibitem{asperti-2012b}
Asperti, A., \emph{A compact proof of decidability for regular expression
  equivalence}, in: L.~Beringer and A.~Felty, editors, \emph{Interactive
  Theorem Proving},  Lecture Notes in Computer Science  \textbf{7406}, Springer
  Berlin Heidelberg, 2012 pp. 283--298.
\newline\urlprefix\url{http://dx.doi.org/10.1007/978-3-642-32347-8_19}

\bibitem{barthwal-norrish-2010a}
Barthwal, A. and M.~Norrish, \emph{A formalisation of the normal forms of
  context-free grammars in {HOL4}}, in: A.~Dawar and H.~Veith, editors,
  \emph{Computer Science Logic, 24th International Workshop, CSL 2010, 19th
  Annual Conference of the EACSL, Brno, Czech Republic, August 23--27, 2010.
  Proceedings},  Lecture Notes in Computer Science  \textbf{6247} (2010), pp.
  95--109.

\bibitem{barthwal-norrish-2010b}
Barthwal, A. and M.~Norrish, \emph{Mechanisation of {PDA} and grammar
  equivalence for context-free languages}, in: A.~Dawar and R.~J. G.~B.
  de~Queiroz, editors, \emph{Logic, Language, Information and Computation, 17th
  International Workshop, {WoLLIC}~2010},  Lecture Notes in Computer Science
  \textbf{6188}, 2010, pp. 125--135.

\bibitem{barthwal-norrish-2013}
Barthwal, A. and M.~Norrish, \emph{A mechanisation of some context-free
  language theory in {HOL4}}, Journal of Computer and System Sciences (WoLLIC
  2010 Special Issue, A. Dawar and R. de Queiroz, eds.) \textbf{80} (2014),
  pp.~346 -- 362.
\newline\urlprefix\url{http://www.sciencedirect.com/science/article/pii/S0022000013000925}

\bibitem{bertot-2004}
Bertot, Y. and P.~Castéran, \enquote{Interactive Theorem Proving and Program
  Development,} Springer, 2004.

\bibitem{braibant-2010}
Braibant, T. and D.~Pous, \emph{An efficient {C}oq tactic for deciding {K}leene
  algebras}, in: M.~Kaufmann and L.~Paulson, editors, \emph{Interactive Theorem
  Proving},  Lecture Notes in Computer Science  \textbf{6172}, Springer Berlin
  Heidelberg, 2010 pp. 163--178.
\newline\urlprefix\url{http://dx.doi.org/10.1007/978-3-642-14052-5_13}

\bibitem{braibant-2012}
Braibant, T. and D.~Pous, \emph{Deciding {K}leene algebras in {C}oq}, Logical
  Methods in Computer Science \textbf{8} (2012), pp.~1--42.
\newline\urlprefix\url{http://arxiv.org/pdf/1105.4537.pdf}

\bibitem{briais-2008}
Briais, S., \emph{Finite automata theory in {C}oq},
  \url{http://sbriais.free.fr/}, [Online; accessed July 28th, 2014].

\bibitem{constable-1997}
Constable, R.~L., P.~B. Jackson, P.~Naumov and J.~Uribe, \emph{Constructively
  formalizing automata theory}, in: \emph{Proof, Language and Interaction:
  Essays in Honour of Robert Milner}, 1997.

\bibitem{coq-2012}
{Coq} {Development}~{Team}, T., \enquote{The {Coq} Reference Manual, version
  8.4,}  (2012), available electronically at \url{http://coq.inria.fr/doc}.

\bibitem{coquand-2011}
Coquand, T. and V.~Siles, \emph{A decision procedure for regular expression
  equivalence in type theory}, in: J.-P. Jouannaud and Z.~Shao, editors,
  \emph{Certified Programs and Proofs},  Lecture Notes in Computer Science
  \textbf{7086}, Springer Berlin Heidelberg, 2011 pp. 119--134.
\newline\urlprefix\url{http://dx.doi.org/10.1007/978-3-642-25379-9_11}

\bibitem{doczkal-2013}
Doczkal, C., J.-O. Kaiser and G.~Smolka, \emph{A constructive theory of regular
  languages in {C}oq}, in: G.~Gonthier and M.~Norrish, editors, \emph{Certified
  Programs and Proofs},  Lecture Notes in Computer Science  \textbf{8307},
  Springer International Publishing, 2013 pp. 82--97.
\newline\urlprefix\url{http://dx.doi.org/10.1007/978-3-319-03545-1_6}

\bibitem{pierce}
et~al., B. C.~P., \emph{{S}oftware {F}oundations},
  \url{http://www.cis.upenn.edu/~bcpierce/sf/current/index.html}, [Online;
  accessed January 23rd, 2015].

\bibitem{filliatre-1997a}
Filliâtre, J.-C., \emph{Finite automata theory in {C}oq: A constructive proof
  of {K}leene's theorem}, Research Report 97--04, LIP - ENS Lyon (1997).
\newline\urlprefix\url{ftp://ftp.ens-lyon.fr/pub/LIP/Rapports/RR/RR97/RR97-04.ps.Z}

\bibitem{firsov-2014}
Firsov, D. and T.~Uustalu, \emph{Certified \{CYK\} parsing of context-free
  languages}, Journal of Logical and Algebraic Methods in Programming
  \textbf{83} (2014), pp.~459 -- 468, the 24th Nordic Workshop on Programming
  Theory (NWPT 2012).
\newline\urlprefix\url{http://www.sciencedirect.com/science/article/pii/S2352220814000601}

\bibitem{firsov-2015}
Firsov, D. and T.~Uustalu, \emph{Certified normalization of context-free
  grammars}, in: \emph{Proceedings of the 2015 Conference on Certified Programs
  and Proofs}, CPP '15 (2015), pp. 167--174.
\newline\urlprefix\url{http://doi.acm.org/10.1145/2676724.2693177}

\bibitem{hopcroft-1979}
Hopcroft, J.~E. and J.~D. Ullman, \enquote{Introduction To Automata Theory,
  Languages and Computation,} Addison-Wesley Publishing Co., Inc., 1979.

\bibitem{coq-site}
INRIA, \emph{The {C}oq proof assistant}, \url{http://coq.inria.fr/}, [Online;
  accessed July 28th, 2014].

\bibitem{jourdan-2012}
Jourdan, J.-H., F.~Pottier and X.~Leroy, \emph{Validating {L}{R}(1) parsers},
  in: \emph{Proceedings of the 21st European Conference on Programming
  Languages and Systems}, ESOP'12 (2012), pp. 397--416.
\newline\urlprefix\url{http://dx.doi.org/10.1007/978-3-642-28869-2_20}

\bibitem{kaloper-1996}
Kaloper, M. and P.~Rudnicki, \emph{Minimization of finite state machines},
  Formalized Mathematics \textbf{5} (1996), pp.~173--184.

\bibitem{koprowski-2010}
Koprowski, A. and H.~Binsztok, \emph{{TRX}: A formally verified parser
  interpreter}, in: \emph{Proceedings of the 19th European Conference on
  Programming Languages and Systems}, ESOP'10 (2010), pp. 345--365.
\newline\urlprefix\url{http://dx.doi.org/10.1007/978-3-642-11957-6_19}

\bibitem{krauss-2012}
Krauss, A. and T.~Nipkow, \emph{Proof pearl: Regular expression equivalence and
  relation algebra}, Journal of Automated Reasoning \textbf{49} (2012),
  pp.~95--106.
\newline\urlprefix\url{http://dx.doi.org/10.1007/s10817-011-9223-4}

\bibitem{kreitz-1986}
Kreitz, C., \emph{Constructive automata theory implemented with the {N}uprl
  proof development system}, Research report, Cornell University, Ithaca, NY
  (1986).

\bibitem{miyamoto}
Miyamoto, T., \emph{{R}eg{E}xp},
  \url{http://coq.inria.fr/pylons/contribs/view/RegExp/trunk}, [Online;
  accessed July 28th, 2014].

\bibitem{moreira-2009}
Moreira, N., D.~Pereira and S.~M. de~Sousa, \emph{On the mechanization of
  {K}leene algebra in {C}oq}, Technical Report DCC-2009-03, DCC-FC-LIACC,
  Universidade do Porto (2009).
\newline\urlprefix\url{http://www.dcc.fc.up.pt/Pubs/}

\bibitem{moreira-2012}
Moreira, N., D.~Pereira and S.~M. de~Sousa, \emph{Deciding regular expressions
  (in-)equivalence in {C}oq}, in: W.~Kahl and T.~Griffin, editors,
  \emph{Relational and Algebraic Methods in Computer Science},  Lecture Notes
  in Computer Science  \textbf{7560}, Springer Berlin Heidelberg, 2012 pp.
  98--113.
\newline\urlprefix\url{http://dx.doi.org/10.1007/978-3-642-33314-9_7}

\bibitem{norrish-2011}
Norrish, M., \emph{Mechanised computability theory}, in: M.~Eekelen,
  H.~Geuvers, J.~Schmaltz and F.~Wiedijk, editors, \emph{Interactive Theorem
  Proving},  Lecture Notes in Computer Science  \textbf{6898}, Springer Berlin
  Heidelberg, 2011 pp. 297--311.
\newline\urlprefix\url{http://dx.doi.org/10.1007/978-3-642-22863-6_22}

\bibitem{ramos-2014}
Ramos, M. V.~M. and R.~J. G.~B. de~Queiroz, \emph{{Formalization of closure
  properties for context-free grammars}}, ArXiv e-prints  (2015).
\newline\urlprefix\url{http://adsabs.harvard.edu/abs/2015arXiv150603428R}

\bibitem{ramos-2009}
Ramos, M. V.~M., J.~J. Neto and I.~S. Vega, \enquote{Linguagens Formais: Teoria
  Modelagem e Implementa\c{c}\~ao,} Bookman, 2009.

\bibitem{ridge-2011}
Ridge, T., \emph{Simple, functional, sound and complete parsing for all
  context-free grammars}, in: \emph{CPP}, 2011, pp. 103--118.

\bibitem{sudkamp-2006}
Sudkamp, T.~A., \enquote{Languages and {M}achines,} Addison-Wesley, 2006, 3rd
  edition.

\bibitem{wu-2011}
Wu, C., X.~Zhang and C.~Urban, \emph{A formalisation of the {M}yhill-{N}erode
  theorem based on regular expressions (proof pearl)}, in: M.~Eekelen,
  H.~Geuvers, J.~Schmaltz and F.~Wiedijk, editors, \emph{Interactive Theorem
  Proving},  Lecture Notes in Computer Science  \textbf{6898}, Springer Berlin
  Heidelberg, 2011 pp. 341--356.
\newline\urlprefix\url{http://dx.doi.org/10.1007/978-3-642-22863-6_25}

\bibitem{xu-2013}
Xu, J., X.~Zhang and C.~Urban, \emph{Mechanising {T}uring {M}achines and
  computability theory in {I}sabelle/{HOL}}, in: S.~Blazy, C.~Paulin-Mohring
  and D.~Pichardie, editors, \emph{Interactive Theorem Proving},  Lecture Notes
  in Computer Science  \textbf{7998}, Springer Berlin Heidelberg, 2013 pp.
  147--162.
\newline\urlprefix\url{http://dx.doi.org/10.1007/978-3-642-39634-2_13}

\end{thebibliography}
\end {document}